\documentclass[12pt,a4paper]{article}
\usepackage{eufrak,bbm,a4,amsmath}
\newcommand{\Bbb}{\mathbbm}
\newcommand{\frak}{\mathfrak}

\def\diag{\mbox{diag}}

\def\Ad{\mbox{Ad}}
\def\ad{\mbox{ad}}

\def\R{\mbox{R}}

\def\d{\mbox{d}}

\def\F{\mbox{F}}
\def\A{\mbox{A}}
\def\tr{\mbox{tr}}

\def\det{\mbox{det}}

\begin{document}

\begin{center}

{\Large Exact  solutions in \\

 Einstein-Yang-Mills-Dirac systems \\}

\vspace{2cm}

{\large Gerd Rudolph and Torsten Tok}

\vspace{1cm}

{ Institute of Theoretical Physics, University Leipzig \\
04109 Leipzig, Augustusplatz 10 \\}

\vspace{1,5cm}

{\large Igor Volobuev}

\vspace{1cm}

{Institute of Nuclear Physics, Moscow State University \\
119899 Moscow, Russia \\}

\end{center}

\begin{abstract}
We present exact solutions in Einstein-Yang-Mills-Dirac theories 
with gauge groups $ SU ( 2 ) $ and $ SU ( 4 ) $ in
Robertson-Walker space-time $ {\Bbb R} \times S^3 $, which are
symmetric under the action 
of the group $ SO ( 4 ) $ of spatial rotations. 
Our approach is based on the dimensional reduction method for gauge 
and gravitational fields and relates symmetric solutions in EYMD theory 
to certain solutions of an effective dynamical system. 

We interpret our solutions as  cosmological solutions with an oscillating  
Yang-Mills field passing between topologically distinct vacua. 
The explicit form of the solution for spinor field shows that its
energy changes the sign during the evolution of the Yang-Mills field 
from one vacuum to the other, which can be considered as 
production or annihilation of fermions.  

Among the obtained solutions there is also a static sphaleron-like
solution, which is a cosmological analogue of the first
Bartnik-McKinnon solution in the presence of fermions. 
\end{abstract}

\section{Introduction}

Exact solutions in gravity coupled to fields of different types have always 
attracted much attention. In particular, the last few years witnessed a great
interest in solutions to EYM systems. There were found both
numerical \cite{B/M,Bizon} and
exact solutions with $SO(3)$ 
and $SO(4)$ groups of spacial symmetry \cite{galtsov,we}. The exact 
$SO(3)$-symmetric solutions turn out to be static and singular and are, 
in fact,
a generalization of Reissner-Nordstr\"om solutions to non-abelian 
gauge theories. The $SO(4)$-symmetric solutions correspond to the 
Robertson-Walker ansatz for the metric and are interpreted either as wormhole 
solutions 
in the euclidean domain \cite{bert}   or as cosmological solutions for the 
radiation
dominated universe \cite{bm}. 

There were attempts to accommodate spinor fields in this picture as well, but 
they usually amounted to considering spinor fields in the EYM background. For
example, in \cite{Gibbons,Gibbons1,V2} 
an anomalous fermion production in the EYM 
background was discussed.

The study of exact solutions in Einstein-Dirac systems also attracted much
attention \cite{Henneaux,Jantzen}. It was found that  the Robertson-Walker
ansatz in such a system leads  to the so-called                                  
"ghost solutions" in ED systems, for which the Dirac field has a vanishing
energy-momentum tensor.

In the present paper we consider a self-consistent EYMD system and find exact
solutions for the case of the gauge groups $SU(2)$ and  $SU(4)$. 
The  group $ SU(4) $ is  the simplest gauge  group which
gives qualitatively new results in comparison with the group 
$ SU(2) $. Our solutions describe the 
radiation dominated universe, in which an exchange of energy 
between the YM and Dirac fields takes place.  They can be of
interest for treating the dynamics of the early universe, because
they are based on the energy-momentum tensor derived from the
fundamental Lagrangians of particle physics, rather than  on the
phenomenological one. Our solutions  also solve the
problem of the "ghost solutions" in ED systems.

To find these solutions we employ the dimensional reduction method
for gravitational, gauge and spinor
fields \cite{Manton,KMRV}, which enables us to relate symmetric solutions in
EYMD system to certain solutions of an effective dynamical system
and essentially simplifies the problem of finding symmetric solutions.

\section{The effective action}

We consider an $ SO(4)$-symmetric Einstein-Yang-Mills-Dirac system 
with the standard action in space-time  $ M = {\Bbb R} \times 
S^3 $: 
\begin{equation}
S = S_E + S_{YM} + S_D \, ,
\end{equation}
where 
\begin{eqnarray}
\label{action_E}
S_E &=& \frac{1}{16 \pi \kappa} \int_M 
\, ( \R - \Lambda ) \, \d v  \, , \\
\label{action_YM}
S_{YM} &=&  \frac{1}{8 g^2} \int_M
\mbox{tr} ({\hat {\F}}_{\alpha \beta}
{\hat {\F}}^{\alpha \beta}) \, \d v \, , \\
\label{action_D}
S_D &=& \int_M 
\frac{i}{2} \bar \psi \gamma^\alpha {E_\alpha}^\mu \left(
\frac{\partial}{\partial x^\mu} - \frac{1}{8} \omega_{\mu \alpha \beta}
[ \gamma^\alpha , \gamma^\beta ] + \delta ' ( \hat {\A}_\mu ) \right) \psi \,
 \d v + h.c. \, .
\end{eqnarray}
Here $ \Lambda $ denotes the cosmological constant, $ \R $ is the 
scalar curvature and $ d v = \sqrt{|\det({\bf g})|} \d^4 x $ 
is the canonical volume form corresponding 
to a metric $ {\bf g} $ on $ M = {\Bbb R} \times S^3 $. 
$ \hat {\F} $ and $ \hat {\A} $ are the Yang-Mills field strength 
and potential respectively, and the trace is taken in the adjoint
representation   of the Lie algebra of the gauge group.    
In the Dirac action,
$ \psi $ denotes the spinor field, $ \{ E_{\alpha} =  {E_\alpha}^\mu                                               
\partial_\mu \} $ is
an orthonormal frame on $M = {\Bbb R} \times S^3 $, $ \{\gamma^\alpha\} $ 
are the gamma matrices and $ \delta $ is the representation
of the gauge group in spinor space. 


We  identify $ S^3 $ with the group manifold of $ SU(2) $. 
Then the 
action $ \sigma $ of $K = SO(4) \equiv (SU(2) \times SU(2)) / 
\{ {\Bbb I}, - {\Bbb I} \} $ on $ SU(2) $ is given by  
\begin{equation}
\sigma ( (k_1,k_2) , x ) = k_1 x {k_2}^{-1} \, , \quad (k_1,k_2) \in SU(2)
\times SU(2) \, , \, x \in SU(2) \, .
\end{equation}
The isotropy group $ H $ at $ x = {\Bbb I}_{SU(2)} $ is isomorphic to 
$ SU ( 2 ) $ and is given by
\begin{equation}
H = \left\{ (k,k) \in K \,, \quad k \in SU(2)  \right\} \, .
\end{equation}
Further, we have the 
reductive decomposition $ {\frak K } = {\frak H} \oplus {\frak M} $
of the Lie algebra $ \frak K $ of $ K $, where
\begin{eqnarray} 
{\frak H} &=& 
\left\{ ( X , X )  \, ; \quad  X \in su(2) \right\} \, , \\
{\frak M} &=& 
\left\{ ( X , - X ) \, ; \quad  X \in su(2) \right\} \,  . 
\end{eqnarray}
Obviously $ S^3 = (SU(2) \times SU(2)) / H $ is a symmetric space, and 
$ SO(4)$-invariance is equivalent to invariance under $ SU(2) \times 
SU(2) $. The isotropy representation of $ H = SU(2) $ in $ {\frak M} $ 
is $ \underline{3} $, i.e. it is isomorphic to the adjoint representation.

Now we have to reduce the action of the original theory due to the
$SO(4)$-symmetry.  We begin with the action of the gravitational field.

The most general form of an $SO(4)$-invariant metric 
$ {\bf g} $ on $ M $ is
\begin{equation}
\label{metric}
{\bf g} = - N^2 {\d t}^2 + a^2 {\d \Omega_{S^3}^{2}} \, ,
\end{equation}
where $ {\d \Omega_{S^3}^2} $ is the 
standard metric on a 3-sphere of radius 2 resp. $ - \frac{1}{2} $ the
Killing metric on $ SU(2) \equiv S^3 $. For the sake of the
future convenience we assume $N$ and $a$ to have the dimension of
length and $t$ and ${\d \Omega_{S^3}^2}$ to be dimensionless.

An orthonormal coframe $ \{ {\theta}^\mu \} $ on $ M $ is given by
\begin{equation}
\label{frame}
\theta^0 = N \d t \, , \quad \theta^i = a \vartheta^i \, 
, \, i=1,2,3 \, ,
\end{equation}
where $ \vartheta^i \sigma_i / ( 2 i ) \equiv \vartheta $ is the canonical 
left invariant 1-form on $ SU(2) $, $\sigma_i$ being the Pauli matrices. 
In what follows we set $ \tau_k = \sigma_k / ( 2 i ) $ and denote
by $ \eta_{\alpha \beta} $ the Minkowski 
metric $ \eta = \diag(-1,1,1,1) $.
It is a matter of simple calculations 
to find the components of the spin connection on M:
\begin{eqnarray}
\nonumber
{\omega_{0 i}} & = & - {\omega_{i 0}} = - \stackrel{.}{a} \frac{1}{a N} 
\delta_{i l} \theta^l \, , \\
\label{spinconnection}
{\omega_{i k}} & = & - \frac{1}{2 a} {\varepsilon_{i k l}} \theta^l \, ,
\quad i = 1 , 2 , 3 \, .   
\end{eqnarray}
Using the standard formulae for the curvature, substituting it
into (\ref{action_E}) and integrating over
$S^3$, one easily finds the reduced gravitational action 
\begin{eqnarray}
S_E &=&  \frac{16 \pi^2}{16 \pi \kappa} \, 
\int_{{\Bbb R}}  a^3 N  \left(  \frac{3}{2 a^2} + 6  
\left(\stackrel{.}{a} \frac{1}{a N}   
\right)^2 -  \stackrel{.}{a} \stackrel{.}{N} \frac{6}{a N^3} 
+ \stackrel{..}{a} \frac{6}{a N^2} - \Lambda \right) \d t \, ,
\end{eqnarray}
where $ 16 \pi^2 $ is the volume of $ SU(2) = S^3 $ with the standard 
metric $ \d {\Omega_{S^3}^2} $.
Omitting a complete divergence, we get the effective action
\begin{equation}
S_E = \frac{16 \pi^2}{16 \pi \kappa} \, \int_{{\Bbb R}} 
\left(  \frac{3}{2} a N - \Lambda a^3 N - 6 \frac{1}{N} a 
{\stackrel{.}{a}}^2 \right) \d t \, ,
\end{equation}
which we consider as the reduced action of the gravitational field.

Next we turn to the gauge field action (\ref{action_YM}).
An $ SO(4)$-symmetric gauge potential $ \hat {\A} $ 
on $ {\Bbb R} \times S^3 $ is
in one-to-one correspondence with a triplet $\{\tau \, , A \, ,
\Phi \}$, where $ \tau $ is a homomorphism  from the
isotropy group $ H $ into the gauge group $ G $
\begin{equation} 
\tau : H  \rightarrow G  \, ,
\end{equation}
$ {\A} $ is  a gauge potential  on $ \Bbb R $ with values in the centralizer 
$ {\frak C}_{{\frak G}} (\tau^, ( {\frak H })) $ of 
$ \tau^, ( {\frak H }) $ in $ {\frak G} = Lie( G ) $ 
and $\Phi$ is a linear mapping
\begin{equation}
\Phi : {\Bbb R} \rightarrow {\frak M}^* \otimes {\frak G}
\end{equation}
with
\begin{equation}
\label{symm-gauge-field-cond}
\Phi \circ \Ad ( h ) = \Ad ( {\tau(h)} ) \circ \Phi 
\,, \quad  \forall h \in H \, .
\end{equation}
Here $ \Ad ( h ) $ is the restriction to $ H $ of the adjoint 
representation of $ K $ applied to $ \frak M $ and 
$ \Ad ( {\tau(h)} ) $ is the restriction to $ \tau ( H ) $ of the 
adjoint representation of $ G $ in 
$ \frak G $.
As we have already mentioned, we will consider gauge groups
$ SU(2) $ and $ SU(4) $. In the first case the centralizer is trivial,
and there is no reduced gauge potential $ {\A} $, whereas in the second case
there can be a nontrivial centralizer. Therefore, the case of the
group $SU(4)$ is more general, and we will carry out all the
calculations for this case and then explain the difference from the
case of $SU(2)$.

Thus, we take the gauge group $ G = SU(4) $ and 
define the homomorphism $ \tau : H \rightarrow G $ by the decomposition 
of the fundamental representation $ \underline{4} $ of SU(4) :
\begin{equation}
\underline{4} \rightarrow ( \underline{2} , \underline{2} ) \, ,
\end{equation}
i.e. we represent the space $ {\Bbb C}^4 $ by the tensor product 
$ {\Bbb C}^2 \otimes {\Bbb C}^2 $ and let the 
fundamental representation of $ \tau( H) $ act on the first factor
and the fundamental representation of the centralizer 
$ { C}_{{ G}} (\tau ( { H })) = SU(2) $ act  on the second factor. 

It is known that the adjoint representation of $ sl(n) $
can be expressed in 
terms of its
fundamental representation by \cite{Dynkin}
\begin{eqnarray}
\ad \, sl(n) &=& \underline{n}^* \tilde \otimes \, \underline{n} \, ,
\end{eqnarray}
where tilde means dropping out a one-dimensional trivial representation 
and  
$ \underline{n}^* $ denotes the contragradient representation, $ t(x)^* =
- t(x)^T $. Therefore we get 
\begin{equation}
\label{decomp}
\ad \, su(4) = (\underline{2},\underline{2})^* \tilde \otimes 
(\underline{2},\underline{2}) \rightarrow (\underline{3},\underline{1}) 
\oplus ( \underline{3},\underline{3}) \oplus 
( \underline{1}, \underline{3} ) \, .
\end{equation}
A basis  in $ su(4) $ adapted  to this decomposition is 
\begin{eqnarray}
\label{mrule}
H_k &=& \tau_k \otimes {\Bbb I} = 
\left( \begin{array}{cc}
        \tau_k & 0 \\
        0 & \tau_k
       \end{array} \right)
\, , \\
P_{k A} &=& \tau_k \otimes \sigma_A  \, , \\
H_A &=& {\Bbb I} \otimes \tau_A \, ,
\end{eqnarray}
where  equation (\ref{mrule}) explains our rule for
evaluating the tensor product of matrices.
It is easy to calculate the commutators of these generators, for instance
\begin{equation}
[ P_{k A} , P_{l B} ] = 
\varepsilon_{klm} H_m \delta_{A B} + \varepsilon_{ABC} H_C \delta_{k l}
\, .
\end{equation}
The constraint (\ref{symm-gauge-field-cond}) means that the mapping
$\Phi$ is an intertwining operator, which intertwines 
the isotropy representation $\underline{3} $ of $ H=SU(2) $ in
${\frak M} $ with the representation (\ref{decomp}). We introduce
basic intertwining operators from $\frak M$ into $\frak G$ defined
by the relations 
\begin{eqnarray}
I((\tau_k, -\tau_k)) &=& H_k \, , \\
I_A ((\tau_k, -\tau_k)) &=& P_{kA} \, .
\end{eqnarray} 
Then for any $X = (X^k \tau_k) \in su(2) $ we have
\begin{eqnarray}
\Phi((X,-X)) &=&  \xi I(X, -X) + \xi^A I_A(X, -X) \\
= X^k ( \xi H_k + \xi^A P_{k A} )&=& X \otimes ( \xi {\Bbb I} +
\xi^A \sigma_A ) \, , 
\end{eqnarray}
where $ \xi $ and $ \xi^A , A = 1,2,3 $ are real valued functions on 
$ {\Bbb R} $. In what follows we denote 
the vector $ (\xi^1,\xi^2,\xi^3 ) $ by $ \vec{\xi} $ and  set $ \hat \xi = 
\xi {\Bbb I} + \xi^A \sigma_A $ and $ \tilde \xi 
= \xi^A \sigma_A $. 
The centralizer $ {\frak C}_{{su(4)}} (\tau' ( {su(2) })) $
of $ \tau' (su(2)) $ in $ {\frak G} = su(4) $ is $ su(2) $ and is 
spanned by the Lie algebra elements 
$ H_A , A= 1, 2, 3 $. The matrix $ \hat \xi $ 
is in the representation $ (\underline{1} +  \underline{3}) $ of 
the centralizer $ \frak C $ and $ i \hat \xi \in u(2) $.

The symmetric gauge potential $ \hat {\A} $ on 
$ {\Bbb R} \times S^3 $ can be easily expressed in terms of
the matrix $ \hat \xi $ and 
the canonical left invariant 1-form $ \vartheta = \vartheta^i
\tau_i $ on $ SU(2) \equiv S^3 $: 
\begin{eqnarray}
\hat {\A} &=& \frac{\xi + 1}{2} H_i \vartheta^i + 
\frac{1}{2} \xi^A P_{k A} \vartheta^k + \A_0^A H_A \d t \nonumber \\
\label{gauge-field-1}
&=& \frac{1}{2} \vartheta \otimes \left( \hat \xi +  {\Bbb I} \right) + 
{\Bbb I} \otimes \A_0 \d t \, .
\end{eqnarray}
Here $ \A_0 $ is a function on $ {\Bbb R} $ with values in 
$ su(2) $, i.e. $ \A_0 = \A_0^A \tau_A $. 
The term $ {\Bbb I} \otimes \A_0 \d t $ 
is the reduced gauge potential on $ \Bbb R $, which can be gauged
out, but we keep it for the moment, because it is necessary for
deriving the equations of motion.
It is not difficult to calculate the corresponding field strength
$ \hat {\F} $ and to obtain the reduced Yang-Mills action, which takes the
simple form 
\begin{eqnarray}
\label{action}
S_{YM} &=&  \frac{16 \pi^2}{8 g^2} \int_{\Bbb R} 24 \Bigg( 
\frac{a}{2 N} \stackrel{.}{\xi}^2 + 
\frac{a}{2 N}  ( \stackrel{.}{\vec{\xi}} + \vec{\A}_0 \times 
\vec{\xi} \, )^2 \nonumber \\
&& - \frac{N}{8 a} \left( ( \xi^2 + (\vec{\xi} \, )^2 -1)^2 + 4 \xi^2 
(\vec{\xi} \, )^2 \right) \Bigg) \d t \, ,
\end{eqnarray} 
where $ \vec{\A}_0 $ is the vector $ (\A_0^1,\A_0^2,\A_0^3) $ and 
$ \vec{\A}_0 \times \vec{\xi} $ is the  vector product
of $ \vec{\A}_0 $ and $ \vec{\xi} $.

If the gauge group is $ G = SU (2) $,  the unique nontrivial homomorphism 
$ \tau : H \rightarrow G $ can be defined by the identity mapping, i.e. 
\begin{equation}
\tau( (k,k) ) = k \,  , \quad   k \in SU(2) \, .
\end{equation}
In this case the centralizer $ {\frak C}_{{\frak G}} (\tau^, (
{\frak H })) $ is trivial, and  the intertwining operator is
\begin{equation}
\Phi( (X,-X) ) = \xi ( t ) X \, , \quad  X \in su(2) \, , \,
\xi(t) \in {\Bbb R} \, .
\end{equation}
It is clear that the gauge potential  $ \hat {\A} $ on  
$ {\Bbb R} \times S^3 $ 
still has the  form (\ref{gauge-field-1}) with $\vec \xi = 0, \A_0 = 0$.
The reduced Yang-Mills action is also given by (\ref{action}),
provided one puts $\vec \xi = 0$ and rescales the coupling
constant $g \rightarrow 2g$.

Next we have to reduce the action (\ref{action_D}) for symmetric
spinor field. We choose $ \delta $ to be the fundamental representation,  
i.e. we can write the spinor $ \psi $  as a $ 4 \times 4 $ 
 resp. $ 4 \times 2 $ 
matrix on which an element $ g $ of 
$ SU(4) $ resp. $ SU(2) $ acts via right multiplication by $ g^{-1} $. 

In accordance with our choice of the metric signature
(\ref{metric}),  we have 
\begin{eqnarray}
\{ \gamma^\mu , \gamma^\nu \} = - 2 \eta^{\mu \nu}\,  , \quad
\eta^{\mu \nu} = diag(- + + + )\, ,  \quad \gamma^5
= i \gamma^0 \gamma^1 \gamma^2 \gamma^3 \, ,\phantom{aaaa} \\ 
\gamma^0  =  \left( \begin{array}{ll}
                   0 & {\Bbb I} \\
                   {\Bbb I} & 0
                   \end{array} \right) \, , \, 
\gamma^i = \left( \begin{array}{ll}
                   0 & -\sigma^i \\
                   \sigma^i & 0
                   \end{array} \right) \, , \,
\gamma^5 = \left( \begin{array}{ll}
                   {\Bbb I} & 0 \\ 
                   0 & -{\Bbb I}
                   \end{array} \right) \, , \, i=1,2,3 \, .
\end{eqnarray}
The bispinor representation $ \Delta ( s ) $ 
is defined by
\begin{equation}
\Delta ( s )^{-1} \gamma^\mu \Delta ( s ) = 
{{\Lambda(s)}^\mu}_\rho \gamma^\rho \, ,
\end{equation}
where $s  $ is an element of the  group $Spin(1,3)$ and $ \Lambda $ is 
the covering homomorphism from $ Spin(1,3) $ onto $ SO(1,3) $.
On Lie algebra level we get 
\begin{equation}
[ \Delta' ( A ) , \gamma^\mu ] = - {A^\mu}_\nu \gamma^\nu \, , \quad 
\Delta' ( A ) = -\frac{1}{8} {A^\mu}_\nu [ \gamma_\mu , \gamma^\nu ] \, ,
\end{equation}
where $ A_{\mu \nu} = - A_{\nu \mu}$ is an element of $ so ( 1,3 )
\equiv spin(1,3) \equiv sl ( 2 , {\Bbb C} ) $.

An $ SO(4) $-symmetric spinor field $ \psi $ on $ {\Bbb R} \times S^3 $ is 
in one-to-one correspondence with a matrix valued function $ \rho
$ on $ {\Bbb R}$,  which satisfies the condition
\begin{equation}
\label{symm-spinor-cond}
\left( \delta ' ( \tau ' ( h )) + \Delta ' ( \lambda ' ( h )) \right) 
\rho = 0 \qquad , \, \forall h \in {\frak H } \, .
\end{equation}
Here $ \lambda' : {\frak H} \rightarrow so(1,3) $ 
is the homomorphism induced by the isotropy representation, which 
can be calculated explicitly:
$ {{\lambda' ( \tau_a )}^b}_c = - {{\varepsilon_{a}}^ b}_ c $.  
Therefore, if the gauge group  $G $ is $ SU( 4) $ equation 
(\ref{symm-spinor-cond}) reads 
\begin{equation}
\label{symm-spinor-cond1}
 \left( \frac{1}{4} \gamma^i \gamma^j \varepsilon_{i j k} 
+ \delta ' \left( H_k \right) \right) \rho =
 \left( \left( \tau_k \otimes {\Bbb I} \right) \rho  
- \rho \left( \tau_k \otimes {\Bbb I} \right) \right) = 0 \, .
\end{equation}
The general solution of this constraint
equation is  
\begin{eqnarray}
\label{symm-spinor}
\rho  = \left( \begin{array}{ll}
                        u_1 {\Bbb I} & u_2 {\Bbb I} \\
                        v_1 {\Bbb I} & v_2 {\Bbb I}
                        \end{array} \right) = {\Bbb I} \otimes 
\left( \begin{array}{ll}
               u_1 & u_2 \\
               v_1 & v_2
               \end{array} \right) 
 \, , \quad u_1 , u_2 , v_1 , v_2 \in C^\infty ( {\Bbb R}) \, ,
\end{eqnarray} 
i.e. a symmetric spinor on $ {\Bbb R} \times S^3 $ is
parameterized by two complex doublets $ u = (u_1 , u_2 )^T $ and 
$ v = (v_1 , v_2 )^T $, 
one for each chirality.  We see from (\ref{symm-spinor}) that the 
reduced gauge group $ C = SU(2) $ acts on 
both doublets by the fundamental representation. 
Taking into account equations (\ref{spinconnection}) and
(\ref{gauge-field-1}),  
it is a matter of simple calculations to get the reduced action 
\begin{eqnarray}
\label{spinor-action}
\nonumber
S_D = 16 \pi^2 \int_{\Bbb R} & a^3 N & \Big( \frac{i}{N} 
( \bar u \stackrel{.}{u} - \stackrel{.}{\bar u} u 
+ \bar v \stackrel{.}{v} - \stackrel{.}{\bar v} v ) \\ 
& & + \frac{1}{N} A_0^B ( \bar u \sigma_B u + \bar v \sigma_B v )
- \frac{3}{2 a} ( \bar u \hat \xi u 
- \bar v \hat \xi v ) \Big) \d t \, .
\end{eqnarray}

If the gauge group $ G $ is $ SU(2) $, equation (\ref{symm-spinor-cond})
reads
\begin{eqnarray}
\label{symm-spinor-cond2}
 \left( \left( \tau_k \otimes {\Bbb I} \right) \rho     
- \rho \,  \tau_k  \right) &=&  0 \, ,
\end{eqnarray}
and we get  
\begin{eqnarray}
\label{symm-spinor2}
\rho  = \left( \begin{array}{l}
                        u {\Bbb I} \\
                        v {\Bbb I} 
                        \end{array} \right) = {\Bbb I} \otimes
\left( \begin{array}{ll}
               u \\
               v
               \end{array} \right)
 \, , \quad u, v \in C^\infty ( {\Bbb R}) \, ,
\end{eqnarray} 
i.e.  a symmetric spinor on $ {\Bbb R} \times S^3 $ for $ G = SU(2) $
depends on two arbitrary complex functions $ u $ and $ v $, one
for each chirality. 
The reduced action has the same form (\ref{spinor-action}), if we
put there $\vec \xi = 0, \A_0 =0$. 

Now we can write down the reduced action of the coupled
EYMD system. In the case of the gauge group $ SU(4)$  it has the form
\begin{eqnarray}
\nonumber
S &=& S_{E} + S_{YM} + S_D  \\
\nonumber
&=&  16 \pi^2 \int_{\Bbb R} \Bigg\{ 
\frac{1}{16 \pi \kappa}
\left(  \frac{3}{2} a N - \Lambda a^3 N - 6 \frac{a}{N}
{\stackrel{.}{a}}^2 \right) \, \\
\nonumber
&& + \frac{24}{8 g^2} \Big( 
\frac{a}{2 N} \stackrel{.}{\xi}^2 + 
\frac{a}{2 N}  ( \stackrel{.}{\vec{\xi}} + \vec{A}_0 \times
\vec{\xi} \, )^2 
- \frac{N}{8 a} ( ( \xi^2 + (\vec{\xi} \, )^2 -1)^2 + 4 \xi^2
(\vec{\xi} \, )^2 ) \Big)  \\
\nonumber
&& + \Big( i a^3
( \bar u \stackrel{.}{u} - \stackrel{.}{\bar u} u 
+ \bar v \stackrel{.}{v} - \stackrel{.}{\bar v} v ) \\
\label{EYMD-action}  
& & + a^3 A_0^B ( \bar u \sigma_B u + \bar v \sigma_B v )
- \frac{3a^2 N}{2} ( \bar u \hat \xi u - \bar v \hat \xi v ) \Big) 
\Bigg\} \d t \, .
\end{eqnarray}
If we choose the gauge group to be $SU(2)$, we have  to put 
$\vec \xi = 0$, $ \A_0 =0$ in this action, to rescale the coupling
constant $ g \rightarrow 2 g $,  and to take into account that the
variables $u$ and $v$ are no longer isospinors, but ordinary
functions.

\section{The field equations and solutions}

Variation of this action with respect to $a \, , \, N \, , \, \xi \, , 
\, \vec \xi \, , \, u \, , \, v $ 
and $ \bar u \, , \, \bar v $ gives us the  field equations. When
taken in the special gauge $ \A_0 = 0 $, $ a= N $ (the latter
condition means that $t$ is now the conformal time), they have
 the form
\begin{eqnarray}
\nonumber
a \frac{\delta}{\delta a} S &=&  \frac{1}{16 \pi \kappa} 
\left( \frac{3}{2} a^2 - 3 \Lambda a^4
- 6 {\stackrel{.}{a}}^2 + 12 a {\stackrel{..}{a}} \right) \\
\nonumber
&& + \frac{3}{2 g^2} \left( 
{\stackrel{.}{\xi}}^2 + ({\stackrel{.}{\vec{\xi}}} \, )^2 + 
\frac{1}{4} \left( ( \xi^2 + ({\vec{\xi}} \, )^2 -1 )^2 + 
4 \xi^2 ({\vec{\xi}} \, )^2 \right) \right) \\ 
\label{Einstein-a}
&& + 3 a^3 \left( i ( \bar u \stackrel{.}{u} - \stackrel{.}{\bar u} u
+ \bar v \stackrel{.}{v} - \stackrel{.}{\bar v} v )
- (\bar u \hat \xi u - \bar v \hat \xi v ) \right) = 0 \, , \\ 
\nonumber
a \frac{\delta}{\delta N} S &=& \frac{1}{16 \pi \kappa} \left( 
\frac{3}{2} a^2 - \Lambda a^4
+ 6 {\stackrel{.}{a}}^2 \right) \\
\nonumber
&& - \frac{3}{2 g^2} \left( {\stackrel{.}{\xi}}^2
+ ({\stackrel{.}{\vec{\xi}}} \, )^2 + 
\frac{1}{4} \left( ( \xi^2 + ({\vec{\xi}} \, )^2 -1 )^2 +
4 \xi^2 ({\vec{\xi}} \, )^2 \right) \right) \\
\label{Einstein-N}
&& - \frac{3}{2}  a^3 (\bar u \hat \xi u - \bar v \hat \xi v ) = 0  
\, , \\
\nonumber
\frac{\delta}{\delta \xi} S &=& - \frac{3}{2 g^2} \left( 
 2 {\stackrel{..}{\xi}}  +  
\left( \xi ( \xi^2 + ({\vec{\xi}} \, )^2 -1 ) +
2 \xi ({\vec{\xi}} \, )^2 \right) \right) \\
\label{Yang-Mills-1}
&& - \frac{3}{2} a^3 (\bar u u - \bar v v ) = 0 \, , \\
\nonumber
\frac{\delta}{\delta \xi^A } S &=& - \frac{3}{g^2} \left( 
 {\stackrel{..}{\xi^A}} + \frac{1}{2}   
\left( \xi^A   ( \xi^2 + ({\vec{\xi}} \, )^2 -1 ) +  
2 \xi^2 \xi^A \, \right) \right) \\
\label{Yang-Mills-2}
&& - \frac{3}{2} a^3 (\bar u \sigma^A  u - \bar v \sigma^A v ) = 0 
\, , \\
\label{Dirac-equation-u}
\frac{1}{2} a^{-\frac{3}{2}} \frac{\delta}{\delta \bar u} S &=& 
i \frac{d}{dt}{a^{\frac{3}{2}}u} 
- \frac{3}{4} \hat \xi {a^{\frac{3}{2}}u} = 0 \, , \nonumber \\
 \frac{1}{2} a^{-\frac{3}{2}} \frac{\delta}{\delta  u} S &=&
i \frac{d}{dt}{a^{\frac{3}{2}} \bar u} 
 + \frac{3}{4}   {a^{\frac{3}{2}} \bar u} \hat \xi = 0 \, , \\
\label{Dirac-equation-v}
\frac{1}{2} a^{-\frac{3}{2}} \frac{\delta}{\delta \bar v} S &=& 
i \frac{d}{dt}{a^{\frac{3}{2}}v} 
+ \frac{3}{4} \hat \xi {a^{\frac{3}{2}}v} = 0 \, , \nonumber \\
\frac{1}{2} a^{-\frac{3}{2}} \frac{\delta}{\delta  v} S &=&
i \frac{d}{dt}{a^{\frac{3}{2}} \bar v} 
 - \frac{3}{4}   {a^{\frac{3}{2}} \bar v} \hat \xi = 0 \, .
\end{eqnarray}
Variation with respect to $\A_0$ gives a constraint
\begin{equation}
\label{Yang-Mills-3}
\frac{\delta S}{\delta \A_0^B } = \frac{3}{g^2} \left( 
\varepsilon_{B C D} \xi^C {\stackrel{.}{\xi^D}} \right) + 
a^3 ( \bar u \sigma_B  u + \bar v \sigma_B v ) = 0 \, ,
\end{equation}
which means that the total isospin of the gauge and the spinor
fields equals zero.

Now we will show that it is possible to find exact solutions to
this system of equations. We begin with the simpler case of the
gauge group $SU(2)$. There is no constraint (\ref{Yang-Mills-3})
in this case, and we also have to drop the equation
(\ref{Yang-Mills-2}), to put $ \vec{\xi} = 0 $ in the others and to
rescale $g \rightarrow 2g$.

The Dirac equations (\ref{Dirac-equation-u}) and 
(\ref{Dirac-equation-v}) give
\begin{equation}
\label{C_u,C_v}
 \bar u u  = \frac{C_u}{a^3} \,, \quad
 \bar v v  = \frac{C_v}{a^3}  
\end{equation}
and
\begin{eqnarray}
\label{C_u}
\bar u \stackrel{.}{u} - \stackrel{.}{\bar u} u &=& - \frac{3 i}{2}
 \xi \bar u  u \, , \\
\label{C_v}
\bar v \stackrel{.}{v} - \stackrel{.}{\bar v} v &=& \frac{3 i}{2} 
 \xi \bar v  v \, ,
\end{eqnarray}
where $ C_u $ and $ C_v $ are arbitrary positive constants. We
will discuss the meaning of these constants later, here we
assume that they are proportional to the number of fermions with
positive resp. negative chirality on $ S^3 $. 

Now equations (\ref{Einstein-a}) and (\ref{Einstein-N}) 
simplify considerably. Their sum gives 
\begin{eqnarray}
  \frac{1}{16 \pi \kappa} \left( 3 a^2 - 4 \Lambda a^4
+ 12 a {\stackrel{..}{a}} \right) = 0 \, .  
\end{eqnarray}
Multiplying this equation by $ {\stackrel{.}{a}} / a $ and
integrating we see that
\begin{equation}
\label{total-energy}
\frac{3}{2} a^2 -  \Lambda a^4 +  6 {\stackrel{.}{a}}^2  = {\bf E} \, ,
\end{equation}
where $ {\bf E} $ is an arbitrary constant which has the meaning of 
the total energy of the system. 
Equation (\ref{total-energy}) is the standard Friedmann equation
for the radiation dominated universe and has a simple analog in
mechanics. We can consider $ a $ as the coordinate of a particle
with mass $ 1 $ and energy $ {\bf E} / 12 $ which moves in a 
potential 
\begin{equation}
\label{W(a)}
W ( a ) = \frac{1}{8} a^2 - \frac{1}{12} \Lambda a^4 \, . 
\end{equation}
If $ \Lambda <
\Lambda_{\bf E} = \frac{9}{16 {\bf E}} $, then the motion will 
be periodical. This means that our solution 
describes a universe which first expands and then contracts, where 
$ a = 0 $ corresponds to a singular metric in the beginning and the end.   
If $ \Lambda \geq \Lambda_{\bf E} $, 
then the solution can be either static or 
can describe an expanding universe. 

Next we turn to the YM equation (\ref{Yang-Mills-1}) (we recall
that we have rescaled $g \rightarrow 2g$ in the case under
consideration). Due to
equation (\ref{C_u,C_v}), it decouples  from the equations for
$u$ and $v$:
\begin{eqnarray}
\label{YM-1-1}
 \frac{3}{8 g^2} \left(  
2 {\stackrel{..}{\xi}} +
\left( \xi ( \xi^2 - 1 ) \right) \right) 
+ \frac{3}{2}  (C_u-C_v) = 0 \, .
\end{eqnarray}
The first integral of this equation is 
\begin{eqnarray}
\label{Einstein-N1}
\frac{3}{8 g^2} \left(
 {\stackrel{.}{\xi}}^2 +  \frac{1}{4} ( \xi^2 -1)^2 \right)
+ \frac{3}{2} \xi ( C_u - C_v ) &=& \frac{{\bf{ E}}}{16 \pi \kappa} \, ,
\end{eqnarray} 
where the integration constant  is due to equations 
(\ref{Einstein-N}) and (\ref{total-energy}).

Equation (\ref{Einstein-N1}) also has an analogue in mechanics. 
A point particle with mass $ 1 $, energy 
$ g^2 {\bf E} / ( 12 \pi \kappa ) $  and 
coordinate $ \xi $  moves in a double-well potential 
\begin{equation}
\label{V(xi)}
 V ( \xi ) =  {\frac{1}{8}} ( \xi^2 -1)^2 
+ 2 g^2 \xi ( C_u - C_v ) \, .
\end{equation}
We can interpret the first term in (\ref{Einstein-N1})
as the energy of the Yang-Mills field and the second term as the 
energy of the Dirac field due to the interaction with the gauge
field. The equation describes 
an  exchange of energy of the two fields, the coupling between the gauge and
the spinor field being proportional to the difference of the numbers of left
and right handed fermions $ C_u - C_v $. 
The exact solution of equation (\ref{Einstein-N1}) is possible in 
terms of elliptic functions of the first kind \cite{Ryshik}.

We consider for instance the case, where the energy 
$ \frac{{\bf E}\, g^2 }{12 \pi \kappa} $ 
of the system is larger 
then the local maximum of the potential 
$ V( \xi ) $. 
Then the system will move between the turning points defined by
the real zeros of the polynomial 
\begin{equation}
\frac{{\bf{ E}}\, g^2 }{12 \pi \kappa} - V(\xi) .
\end{equation}
In the case under consideration we have two real zeros $ \alpha $
and $ \beta $, 
$ \alpha > 0 > \beta $, and two complex conjugated zeros 
$ m - i n $ and $ m + i n $, i.e.
\begin{equation}
\frac{1}{8} (\alpha - \xi ) ( \xi -\beta) ( (\xi - m )^2 + n^2 ) = 
\frac{{\bf{ E}}\, g^2 }{12 \pi \kappa} - V(\xi) = \frac{1}{2} 
{{\stackrel{.}{\xi}}}^2 \, .
\end{equation}
We have to solve the integral
\begin{equation}
t ( \xi ) = \int_{\beta}^{\xi} \frac{d x}{ \sqrt{\frac{{\bf E}\, g^2
}{6 \pi \kappa} 
- \frac{1}{4} ( x^2 -1)^2 - 4 g^2 x ( C_u - C_v )}} \, .
\end{equation}
Its exact solution is given by \cite{Ryshik} 
\begin{equation}
t(\xi) = \frac{2}{\sqrt{{pq}}} 
F \left( 2 {\rm arcctg} \sqrt{\frac{q(\alpha-\xi)}{p(\xi-\beta)}} \, , \,
\frac{1}{2} \sqrt{\frac{-(p-q)^2+(\alpha-\beta)^2}{p q}}
\right) \, ,
\end{equation}
where F is the elliptic integral of the first kind and $ p $, $ q $ are 
defined by
\begin{eqnarray}
p^2 = ( m-\alpha)^2 + n^2 &,& q^2 = ( m-\beta)^2 + n^2 \, .
\end{eqnarray}
This function can be inverted, and we can get $\xi(t)$ expressed
in terms of the Jacobi elliptic function $ sinus \,\, amplitudinis $.

With a given solution $ \xi $ we can solve the Dirac equations
(\ref{Dirac-equation-u}) and (\ref{Dirac-equation-v})
\begin{eqnarray}
\label{loesung-spinor-u}
u &=& \sqrt{\frac{C_u}{a^3}} \exp{ \left\{ 
-i \int \frac{3}{4} \xi \d t \right\} } \, , \\
\label{loesung-spinor-v}
v &=& \sqrt{\frac{C_v}{a^3}} \exp{ \left\{ 
i \int \frac{3}{4} \xi \d t \right\} } \, .  
\end{eqnarray}
One easily checks that our solution fulfills the whole system of field 
equations.

Now we consider the case of  gauge group $SU(4)$. The Dirac
equations (\ref{Dirac-equation-u}) and (\ref{Dirac-equation-v}) 
again lead  to (\ref{C_u,C_v}), but instead of (\ref{C_u}) and 
(\ref{C_v}) we now have
\begin{eqnarray}
\bar u \stackrel{.}{u} - \stackrel{.}{\bar u} u &=& - \frac{3 i}{2}
\bar u \hat \xi u  \, , \\
\bar v \stackrel{.}{v} - \stackrel{.}{\bar v} v &=& \frac{3 i}{2} 
\bar v \hat \xi v \, .
\end{eqnarray}
These equations are also sufficient to decouple the Friedmann equation
from the Yang-Mills-Dirac equations, and we get again the equation
(\ref{total-energy}) for the scale factor $ a $.

Now we have to solve the Yang-Mills equations (\ref{Yang-Mills-1})
and (\ref{Yang-Mills-2}). We start with the discussion of the
constraint (\ref{Yang-Mills-3}).
In what follows we restrict ourselves to the case, where
\begin{equation}
\label{assumption-1}
\varepsilon_{B C D} \xi^C {\stackrel{.}{\xi^D}} = 0 \quad \mbox{for} 
\quad B = 1,2,3 \, ,
\end{equation}
i.e.  the  isospins of the gauge field and the Dirac field
vanish separately.
This equation means that the angular momentum of the motion in the
$\vec \xi$-space equals zero, that is the motion goes along a
straight line passing through the origin, and the vector
$\vec{\xi} $ is always proportional to a fixed vector $ \vec{\xi_0} $.
We use the remaining gauge freedom to choose $ \vec{\xi_0} = 
(0 , 0 , 1 ) $ , i.e. $ \tilde \xi = 
\zeta \sigma_3 , \zeta \in {\Bbb R} $.
In this gauge we obtain from the Yang-Mills equations (\ref{Yang-Mills-2}) 
\begin{equation}
\label{constr-1}
\bar u \sigma_A u - \bar v \sigma_A v = 0 \quad \mbox{for} \quad 
A = 1,2 \, .
\end{equation}
Further we get from equations (\ref{Yang-Mills-3}) and (\ref{assumption-1})
\begin{equation}
\label{constr-2}
\bar u \sigma_B  u + \bar v \sigma_B v = 0 \, , \quad B = 1,2,3 \, .
\end{equation}
Hence, we have $ \bar u \sigma_1 u = \bar u \sigma_2 u = 0 $.
These equations have two solutions:
\begin{eqnarray}
\mbox{1st  case:} \, \, && u = a^{-3/2} \alpha_1 w_+ \, ,  
v = a^{-3/2} \beta_1 w_-  \\
\mbox{2nd case:} \, \, && u = a^{-3/2} \alpha_2 w_- \, ,  
v = a^{-3/2} \beta_2 w_+ \, , \\
\mbox{where} \quad && w_+ = \left( \begin{array}{l}   
              1 \\
              0
           \end{array} \right) \, , \,
w_- =  \left( \begin{array}{l} 
              0 \\
              1
           \end{array} \right)  
\end{eqnarray}  
and $ \alpha_i, \beta_i \in {\Bbb C} $, $ |\alpha_i| = |\beta_i| , 
i=1,2 $. In particular, we obtain $ C_u = C_v = |\alpha_i|^2 $ 
in the i'th case. 
If $ \tilde \xi = \zeta \sigma_3 \equiv 0 $, then the Yang-Mills equations 
(\ref{Yang-Mills-2})  demand $ u = v = 0 $. Therefore, we 
consider only the nontrivial case, when $ \tilde \xi \not\equiv 0 $.
Hence we have  two Yang-Mills equations 
(\ref{Yang-Mills-1}) and (\ref{Yang-Mills-2}), which now take the form:
\begin{eqnarray}
\label{YM-1-2}
 \frac{3}{2 g^2} \left(  
2 {\stackrel{..}{\xi}} +
\left( \xi ( \xi^2 + {\zeta}^2 -1 ) +
2 \xi {\zeta}^2 \right) \right)  &=& 0 \, ,\\
\label{YM-2-2}
\frac{3}{2 g^2} \left(
2 {\stackrel{..}{{\zeta}}} +
\left( {\zeta}   ( \xi^2 + {\zeta}^2 -1 ) +  
2 \xi^2 {\zeta}  \right) \right) + 3 S  &=& 0 \, ,
\end{eqnarray}
with $ S := C_u = | \alpha_1 |^2 $ in the first and  $ S := - C_u 
= - | \alpha_2 |^2 $ in the second case.

To solve equations (\ref{YM-1-2}) and (\ref{YM-2-2}), we pass
to new variables in accordance with
\begin{eqnarray}
\label{substitution}
\xi = \frac{1}{{2}} (x+y) \nonumber \, , \\
\zeta = \frac{1}{{2}} (x-y) \, .
\end{eqnarray}
It is easy to check that the equations for $x$ and $y$ decouple
and take the form
\begin{eqnarray}
\label{x}
 \frac{3}{2 g^2} \left(
 2 {\stackrel{..}{x}} +
 x ( x^2  - 1 ) \right) + 3 S  &=& 0
\, , \\
\label{y}
 \frac{3}{2 g^2} \left(
 2  {\stackrel{..}{y}} +
y ( y^2 - 1 ) \right) - 3 S &=& 0
\, .
\end{eqnarray}
The first integrals of these equations  are
\begin{eqnarray}
\label{E_1}
\frac{3}{2 g^2} \left(
 {\stackrel{.}{x}}^2 +  \frac{1}{4} ( x^2 - 1 )^2 \right)
+ 3 S x  &=& \frac{ {\bf E}_1}{8 \pi \kappa} \, , \\
\label{E_2}
\frac{3}{2 g^2} \left(
 {\stackrel{.}{y}}^2 +  \frac{1}{4} ( y^2 - 1 )^2 \right)
- 3 S y  &=& \frac{ {\bf E}_2}{8 \pi \kappa} \, ,
\end{eqnarray}
where due to (\ref{Einstein-N}) and (\ref{total-energy}) the constants 
${\bf E}$ fulfill
\begin{equation}
{\bf E}_1 + {\bf E}_2 = {\bf E} \, .
\end{equation}
These equations can also be solved exactly in terms of Jacobi
elliptic functions, but unlike the case of the group $SU(2)$,
there will be two different periods of motion in $x$ and $y$.

We would like to note here that equations of the type
(\ref{YM-1-2}), (\ref{YM-2-2}) with $S = 0$ for the Euclidean EYM
system were first found in \cite{bert}, but they were solved there
only for the case either $\xi = 0$ or $\zeta = 0$.

Substituting the solutions for $\xi$ and $\zeta$ into equation
(\ref{gauge-field-1}) we get
\begin{equation}
\label{su(2)+su(2)-zerlegung}
\hat {\A} = \frac{1}{2} \theta \otimes \left( \hat \xi +  
{\Bbb I} \right) =
\left( \begin{array}{ll}
             \frac{1+x}{2} \,   \theta& 0 \\
              0 &  \frac{1+y}{2} \, \theta
       \end{array} \right) \, ,
\end{equation}
i.e.  the gauge potential
 $ \hat {\A} $ takes values only in an $ su(2) \oplus su(2) $ subalgebra of $
su ( 4 ) $, each $ su ( 2 ) $ part of the gauge potential being
coupled to only one of the spinor fields $ u $ resp. $ v $.

If we have a solution to the system of equations 
(\ref{E_1}) and (\ref{E_2}) it is easy to integrate 
the Dirac equations (\ref{Dirac-equation-u}) and 
(\ref{Dirac-equation-v}). With given solutions $ x $ and $
y $  we obtain 
\begin{eqnarray}
\label{loesung-spinor-u-fall1}
u &=& w_+ \sqrt{\frac{C_u}{a^3}} \exp{\left\{
-i \int \frac{3}{4} \, x \, \d t \right\}} \, , \\
\label{loesung-spinor-v-fall1}
v &=& w_- \sqrt{\frac{C_u}{a^3}} \exp{\left\{
i \int \frac{3}{4} \, y \, \d t \right\}} 
\end{eqnarray}
in the first and 
\begin{eqnarray}
\label{loesung-spinor-u-fall2}
u &=& w_- \sqrt{\frac{C_u}{a^3}} \exp{\left\{
-i \int \frac{3}{4} \, y \, \d t \right\}} \, , \\
\label{loesung-spinor-v-fall2}
v &=& w_+ \sqrt{\frac{C_u}{a^3}} \exp{\left\{
i \int \frac{3}{4} \, x \, \d t \right\}} 
\end{eqnarray}
in the second case.

\section{Discussion}

In studying the self-consistent EYMD system we found that the
evolution of the metric decouples from the remaining system and 
is described by the Friedmann equation for 
the radiation dominated universe.
The same result for the case of EYM systems was obtained earlier
in \cite{bert,bm}, and solutions for the spinor field in this
background were studied in \cite{Gibbons1,Molnar}.  Unlike the
latter solutions, our solutions take into account the back
reaction of the spinor field on the YM field.

The Yang-Mills equations in the case $ G = SU(2) $ admit three
static solutions: two minima and one local maximum of the potential
\begin{equation}
V ( \xi ) =
\frac{1}{8} ( \xi^2 - 1 )^2 + 2 g^2 \xi ( C_u - C_v ) \, .
\end{equation}
Of course, these are solutions of the whole system only if the constant
$ {\bf E} $ is  equal to $ V ( \xi ) $.
If we have $ C_u = C_v $, we can interpret these extrema as two vacua
($ \xi = -1 \, , \, \xi = +1 $) with Chern-Simons numbers $ 0 $ and $ 1 $
and as a sphaleron-like solution ($\xi = 0 $) with Chern-Simons number
$ 1/2  $ lying on top of the potential barrier between the vacua
\cite{Molnar,Gibbons1}.    
Chiral spinor fields shift slightly the location of the extrema
and the Chern-Simons index of the corresponding gauge field
configurations, which is quite natural in the presence of matter
fields \cite{Christ,Hellmund,Rubakov}.

By fine tuning the cosmological constant $ \Lambda $ and the energy
$ {\bf E} $, we can get a static sphaleron-like solution of the whole system.
This solution corresponds to the local maxima of
$ V ( \xi ) $ and $ W ( a ) $, see
equations (\ref{V(xi)}) and (\ref{W(a)}).    
Obviously, this solution has two unstable    
modes -- one in the gravitational and one in the gauge field sector.
This is another indication that the static solution is a cosmological
analog of the first Bartnik-McKinnon solution 
\cite{Straumann-96,BHLSV1,BHLSV2}.

At this point, we have to comment on the meaning of our
classical spinor field. The problem of interpreting spinor fields
in cosmology has been discussed for many
years \cite{Henneaux,Jantzen}, but  so far no satisfactory solution has
been found. It is clear that our solution describes just one
energy level of the Dirac field in the EYM background, which is
exactly the so called zero mode. If we assume that the Dirac field
is normalized to unity, the influence of this field on the EYM
system is negligible. Therefore, we suggest that the Dirac field of
our solution is normalized arbitrarily and is, in fact, an
effective field describing fermionic matter with $SO(4)$-invariant
energy-momentum tensor. This assumptions seems to be reasonable in
the cosmological setting, because the energy levels of the Dirac
field in the EYM background must be very dense, and replacing the
contribution of fermions on the lowest levels by that of the zero
mode level could be a good approximation.

Equation (\ref{Einstein-N}) is the $ (0,0) $ component of the Einstein 
equations, and therefore we can interpret the constant $ {\bf E} $
as the total energy of the  YMD system. 
On the other hand, equations (\ref{Einstein-N1}) resp. 
(\ref{E_1}) and (\ref{E_2}) describe the exchange of energy between 
Yang-Mills and spinor field. This interpretation is supported by the 
solutions (\ref{loesung-spinor-u}) and 
(\ref{loesung-spinor-v})
for the spinor field in the case of the gauge group $SU(2)$: the
momentary frequency resp. energy of the spinor 
field is given by the integrand in the exponent of the solutions and this 
is up to a factor $ \xi $.

This observation also means that our solutions describe creation
and annihilation of fermions.
If the total energy of the YMD system is larger than the local maximum
of the potential $ V(\xi) $, the motion in the variable $\xi$,
stemming from the YM field, will be periodical. When 
 $ \xi $ crosses the maximum of the potential $ V $, 
it  changes its sign. With the above interpretation we see that 
the energy of the spinor field also changes its sign.
This is in accordance with the observations in 
\cite{Gibbons1,V2,V3}, where it  was shown that the spinor field has zero modes 
in the sphaleron-background and that moving between neighbouring 
vacua of the gauge field results in a shift of the energy level of the 
spinor field.  We also obtain from our solution that the effect  is opposite 
for the spinor field with opposite chirality. The corresponding
violation of the fermion number can be calculated explicitly.

The classical $ U(1) $ vector current of the Dirac field is given by
\begin{equation}
j = \bar \psi \gamma_\alpha \psi \theta^\alpha 
= \frac{C_u + C_v}{a^3} \theta^0 \, .
\end{equation}
This current is 
classically conserved, i.e. $ d ( \ast j ) = 0 $. 
Here the star denotes the Hodge star
and $ \{\theta^\mu\} $ is the orthonormal coframe on
$ {\Bbb R} \times S^3 $, see equation (\ref{frame}).
But if one considers chiral matter on 
quantum level, this current has an anomaly \cite{Morosov,Rubakov}: 
\begin{equation}
\label{anomaly}
d(\ast j) = (1/16\pi^2) \tr (\hat {\F} \wedge \hat {\F}) \, .  
\end{equation}
Integrating (\ref{anomaly}) over $ I \times S^3 $, $ I = [ t_i , t_f ] $, 
we get 
\begin{equation}
\label{diff-top-charge}
N_F ( t_f ) - N_F ( t_i ) = (1/16\pi^2) \int_{I \times S^3} 
\tr (\hat {\F} \wedge \hat {\F}) = \int_{S^3} Q ( t_f ) 
- \int_{S^3} Q ( t_i ) \, ,
\end{equation}
where $ Q $ is the Chern-Simons 3-form 
\begin{equation}
Q = \frac{1}{16 \pi^2}\tr( \hat {\A} \wedge d \hat {\A} + \frac{1}{3}
\hat {\A} \wedge [ \hat {\A}, \hat {\A}]) \, , 
\end{equation}
and 
\begin{equation} 
N_F ( t ) = \int_{\{t\} \times S^3} \ast j   
\end{equation}
is the total fermion number at time $ t $. 
Hence, for chiral fermions one obtains on quantum level that the change 
in the fermion number 
is given by the difference of the topological charges of the gauge field 
at the two ends of the time interval $ I $. 
The topological charge 
$$ q = \int_{S^3} Q$$
at any fixed $t$ can be calculated from the formula for the symmetric
potential $ \hat{ \A }$, which gives
$$  q(t)= \frac{1}{4}( 2 + 3 \xi(t) - \xi(t)^3). $$
If we pass from the configuration with $ \xi = -1 $ at $ t = t_i $ to 
the configuration with $ \xi = 1 $ at $ t = t_f $, 
the topological charge changes by $ 1 $. 
Setting in our solution one of the constants, $ C_u $ or $ C_v $, equal
to zero, which implies that we have a chiral theory with either
left handed or right handed fermions from the very beginning, we
can interpret the corresponding energy level crossing 
of the Dirac field as creation or annihilation of 
fermions. We would like to emphasize that this process
takes place in real time, in contrast to instanton like effects 
related to barrier penetration in Euclidean space time.

Finally, we shortly comment on the case of gauge group $ SU(4) $.  
Under assumption (\ref{assumption-1}), we can completely 
solve the field equations. We get two spinor fields with opposite chirality 
and equal density ($ C_u = C_v $), see equations 
(\ref{loesung-spinor-u-fall1}) and
(\ref{loesung-spinor-v-fall1}), resp.
(\ref{loesung-spinor-u-fall2}) and
(\ref{loesung-spinor-v-fall2}).

Comparing them to 
(\ref{su(2)+su(2)-zerlegung}) 
we find  that the Yang-Mills potential takes  values only in an 
$ su(2) \oplus su(2) $ subalgebra of $ su(4) $ and, therefore, 
splits into two parts. 
Each part is coupled to one spinor field of a definite chirality. 
Thus, in some sense, we simply have a doubling  of the solution for 
$ SU(2) $.

As it was mentioned above, equations (\ref{E_1}) and 
(\ref{E_2}) describe the exchange of energy between spinor and gauge field. 
Therefore, in the case of gauge group $ SU(4) $ we also have   
energy-level crossing in the evolution of the spinor field. 
But, in contrast to the $ SU(2) $-case, we 
have - due to constraints (\ref{constr-1}) and (\ref{constr-2}) - 
only solutions with equal density of left and right handed fermions, 
i.e. we have no chiral solutions.

\section*{Acknowledgements}

The authors are grateful to Yu.~Kubyshin for fruitful discussions.
One of the authors (I.V.) is grateful to the Center of Natural Sciences of
the University of Leipzig for the warm hospitality extended to him during
his stay in Leipzig. He also acknowledges partial support under the
INTAS-93-1630-EXT project.


\end{document}